\documentclass[aps,twocolumn,showpacs,prl,preprintnumbers,amsmath,amssymb,superscriptaddress]{revtex4-2}
\usepackage{epsf}
\usepackage{graphicx}
\usepackage{sidecap}
 \usepackage{soul}
 \usepackage{array}
 \usepackage{amsmath}
 \usepackage{amssymb}
 \usepackage{dcolumn}
 \usepackage{epstopdf}
 \usepackage{bm}
 \usepackage{amsmath}
 
\usepackage{gensymb}
\usepackage{color}
\usepackage{hyperref}
\usepackage{soul}
\sethlcolor{green}
\usepackage{bbding}
\usepackage{sidecap}
\usepackage{float}

\hypersetup{
    colorlinks=true,
    citecolor=red,
    linkcolor=red,
    filecolor=blue,   
    urlcolor=blue,
}
 
\def \beq {\begin{equation}}
\def \eeq {\end{equation}}
\renewcommand{\figurename}{\textbf{Fig.}}
\renewcommand{\thefigure}{{\textbf{\arabic{figure}}}}

\def\bibsection{\refname}
\renewcommand{\refname}{\noindent\textbf{References}}
\begin{document}
\title{Observation of multiple flat bands and Van-Hove singularities in the distorted kagome metal NdTi$_3$Bi$_4$}

\author{Mazharul Islam Mondal}  \thanks{These authors contributed equally in this work.} \affiliation{Department of Physics, University of Central Florida, Orlando, Florida 32816, USA} 
\author{Anup Pradhan Sakhya}  \thanks{These authors contributed equally in this work.} \affiliation {Department of Physics, University of Central Florida, Orlando, Florida 32816, USA} 
\author{Milo Sprague} \affiliation{Department of Physics, University of Central Florida, Orlando, Florida 32816, USA} 
\author{Brenden R. Ortiz} \affiliation {Materials Science and Technology Division, Oak Ridge National Laboratory, Oak Ridge, Tennessee 37830, USA}
\author{Matthew Matzelle} \affiliation{Department of Physics, Northeastern University, Boston, Massachusetts 02115, USA}
\author{Arun K Kumay} \affiliation{Department of Physics, University of Central Florida, Orlando, Florida 32816, USA} 
\author{Avike Seal} \affiliation{Department of Physics, University of Central Florida, Orlando, Florida 32816, USA} 
\author{Barun Ghosh} \affiliation{Department of Physics, Northeastern University, Boston, Massachusetts 02115, USA}  \affiliation{Department of Condensed Matter and Material Physics, S.N. Bose National Center for Basic Sciences, Kolkata, 700106, India}
\author{Arun Bansil} \affiliation{Department of Physics, Northeastern University, Boston, Massachusetts 02115, USA}
\author{Madhab Neupane} \thanks{Corresponding author:\href{mailto:madhab.neupane@ucf.edu}{madhab.neupane@ucf.edu}}\affiliation{Department of Physics, University of Central Florida, Orlando, Florida 32816, USA}

\date{\today}

\begin{abstract}
\indent Kagome materials have attracted enormous research interest recently owing to their diverse topological phases and manifestation of electronic correlation. Here, we present the electronic structure of a distorted ferromagnetic kagome metal, NdTi$_3$Bi$_4$, exhibiting a transition temperature of 9 K. Our investigation employs a combination of angle-resolved photoemission spectroscopy (ARPES) measurements and density functional theory (DFT) calculations. We discover the presence of two ``flat" bands which are found to originate from the kagome structure formed by Ti atoms with major contribution from Ti \textit{d$_{xy}$} and Ti \textit{$d_{x^{2}-y^{2}}$} orbitals. We also observed multiple van Hove singularities (VHSs) in its electronic structure, with one VHS lying near the Fermi level. The ARPES data reveals the existence of Dirac cone at the $\overline{\text{K}}$ point, a finding which is corroborated by our DFT calculations. These findings present detailed electronic structure capable of hosting correlation-driven phenomenon in this novel ferromagnetic kagome metal.
\end{abstract}

\maketitle

 \indent The kagome structural motif, characterized by a two-dimensional honeycomb network of corner-sharing triangles, has proven to be a valuable platform for investigating  complex quantum interactions involving frustrated geometry, topological phenomena, spin, and electronic correlations \cite{Strecka, Guo, Yin, Kang}. Kagome structures are known to exhibit numerous long-range orders with Fermi surface instabilities \cite {Wang}, topological properties and van Hove singularities (VHSs) \cite {Kang, Ye, Yin}. The discovery of kagome superconductors in the AV$_3$Sb$_5$ family has further accelerated research in kagome metals \cite{Ortiz1, Ortiz2, Ortiz3, Lei}. These materials were explored enormously as they exhibit geometric frustration, superconductivity, unconventional charge density wave (CDW) order  \cite{Zhangcdw, Liang}, pair density wave \cite{Chenpdw}, electronic nematicity \cite{Nienem}, and giant anomalous Hall effect \cite{yanggwh}. An analogous family, the Ti-based ATi$_3$Bi$_5$ compounds, was also discovered with potential superconductivity \cite{Yang, GSu}, electronic nematicity \cite{YangTi, LiTi}, and nontrivial electronic structure \cite{Seshadri, Yang}.\\ 
\indent Recently, a new class of materials with AM$_3$X$_4$ (A: Lanthanide, Ca, M:V, Ti, X:Sb, Bi) stoichiometry has been reported \cite{Ortiz5, alexander, alexander1}. In contrast to previously known kagome compounds, AM$_3$X$_4$ possesses a slight orthorhombic distortion which consists of a slight shortening of one triangular bond and a buckling of the kagome plane into a zig-zagged configuration. In addition to this structural distortion, the AM$_3$X$_4$ family features easily replaceable lanthanide elements. The combination of these two influences produces a wider variety of physical characteristics within members of the AM$_3$X$_4$ family. YbV$_3$Sb$_4$, for instance is non-magnetic and shows no distinct thermodynamic transition between 300 K and 60 mK. However when Yb is replaced by Eu, EuV$_3$Sb$_4$ undergoes a ferromagnetic transition below 32 K \cite{Ortiz5}. A pronounced magnetic anisotropy and notable band evolution particularly band splitting are observed during the ferromagnetic phase transition in the titanium-based kagome ferromagnet SmTi$_3$Bi$_4$ \cite{Mahzuza}. These materials feature zig-zag chains of A-site ions and distorted kagome sublattices based on M atoms. Recently, a large number of Ti based $Ln$M$_3$X$_4$ compounds have been synthesized \cite{BrendenLn134}. The key advantage of this new family is the potential to introduce magnetism by selecting a suitable $Ln$ site element analogous to the \textit{R}Mn$_6$Sn$_6$ family \cite{Cochran, Liu, Asaba, Y16, Ma, Gao, WMa, MLi, Dhakal, QWang, Zeng, Gu, Kabir, RSLi, Lv, P}. Consequently, $Ln$M$_3$X$_4$ family offers a promising platform to explore the interplay between magnetism and inherent geometric frustration. ARPES and DFT calculations have shown that $Ln$Ti$_3$Bi$_4$ host Dirac cone and correlated states in its electronic structure \cite{anup, La, Ce}. On the other hand, TbTi$_3$Bi$_4$ exhibits an anisotropic electronic structure where momentum dependent band folding has been observed in its antiferromagnetic (AFM) state \cite{Nishat}. NdTi$_3$Bi$_4$ is ferromagnetic, with a transition temperature of 9 K \cite{BrendenLn134}. It behaves like a soft ferromagnet with its easy-axis of magnetization lying along the [010] plane. Beyond this report, there has been no detailed study of the electronic structure of this material \cite{BrendenLn134}. A thorough investigation of the electronic band structure will be essential in order to gain deeper insight into the unique electronic properties such as strong correlations, non-trivial band topology, and the effects of structural distortion.\\

\begin{figure*}
	\centering
	\includegraphics[width=\linewidth]{Fig1.png}
	\caption{Crystal structure and electronic structure calculation of NdTi$_{3}$Bi$_{4}$. (a) The NdTi$_3$Bi$_4$ unit cell with Nd, Ti, and Bi atoms represented as gold, blue, and purple balls, respectively. The zig-zag chains are formed by Nd atoms. (b) Unit cell viewed along the $c$-axis. The Ti-based kagome structure is highlighted along this orientation. (c) 
    The orthorhombic bulk Brillouin zone (BZ) is illustrated, highlighting the $\Gamma$, X, Y, A, and Y$^\prime$ high symmetry lines. Above the bulk BZ, the (001) surface-projected pseudo-hexagonal BZ is depicted, also indicating high-symmetry points. (d) Band dispersion from DFT based calculations with $d_{xy/x^2}$, $d_{z^2}$, and $d_{xz/yz}$, orbital projections indicated in red, blue, and green colors respectively. The red, pink, and black arrows highlight the presence of VHSs, Dirac-like states, and flat bands, respectively. 
	\label{fig:fig}
    }
\end{figure*}

\indent In this Letter, we investigate the electronic band structure of the ferromagnetic kagome metal NdTi$_3$Bi$_4$. Our findings reveal the presence of two flat bands and multiple VHSs in its electronic band structure. These flat bands originate from Ti kagome lattice and consist predominantly of Ti \textit{d$_{xy}$} and Ti \textit{$d_{x^{2}-y^{2}}$} orbitals. Furthermore, we detect a Dirac cone at the $\overline{\text{K}}$ surface high-symmetry point, located near the Fermi level (E$_F$). 

NdTi$_3$Bi$_4$ single crystals were grown using a bismuth self-flux method. Synchrotron-based ARPES experiments were conducted at the ALS beamline 10.0.1.1. First-principles electronic structure calculations were performed within the DFT formalism, implemented in the Vienna ab initio simulation package \cite{Kresse1, Kresse2, Kresse3, Kresse4}. For detailed information on the single-crystal growth technique, ARPES measurements, and DFT calculations, refer to the Supplemental Material \cite{SM}. NdTi$_3$Bi$_4$ crystallizes in the orthorhombic space group $Fmmm$ (space group No. 69), similar with other reported members of the $Ln$Ti$_3$Bi$_4$ family \cite{BrendenLn134}. 
The unit cell consists of Ti kagome layers, shown in blue within Fig 1.(a), separated by alternations of Bi (purple balls) and Bi/Nd (gold) zig-zag layers. As illustrated in Fig. 1 (a) and Fig. S1 (a, b), the crystal structure features an alternating stacking of NdBi, TiBi, and Bi layers along the c-axis. When viewed along the c-axis, the zig-zag chains formed by Nd atoms become visible, along with a kagome-like lattice formed by Ti atoms, and both a honeycomb-like and triangular lattice formed by Bi atoms (see Supplemental Material Fig. S1) \cite{SM}. A top-down view of the Ti kagome layer along with the Bi intermediate layer is shown in Fig. 1(b) to better illustrate the relevant in-plane lattice characteristics. 
The quasi-1D chains of rare-earth necessitate the reduction in symmetry from C$_6$ to C$_2$, though the kagome lattice is only slightly distorted from the ideal geometry \cite{BrendenLn134}. Fig. 1(c) displays the bulk Brillouin zone (BZ) of NdTi$_3$Bi$_4$ along with its (001) surface projection. High symmetry points are labeled in accordance with the quasi 6-fold rotational symmetry, however the slight C$_2$ distortion splits the six $\overline{\text{M}}$ and $\overline{\text{K}}$ high-symmetry points of the hexagonal BZ into four $\overline{\text{M}}'$ and $\overline{\text{K}}'$ points, along with two $\overline{\text{M}}$ and $\overline{\text{K}}$ inequivalent surface high symmetry points along the $\text{C}_2$ rotational axis. The electronic band dispersions calculated from bulk DFT-based first-principles calculations, including spin-orbit coupling, are presented in Fig. 1(d). The metallic nature of this sample is recognized by the high number of bands that cross the E$_F$. Among these are two parabolic bands featuring electron-like curvature at the $\Gamma$ point, which extends upward to form apparent van-Hove singularities (VHSs) at the Y and Y$'$-points (see red arrows in Fig. 1(d)). The VHSs at Y and Y$'$ are located at different binding energies, highlighting the influence of C$_6$ symmetry-breaking. Similarly, we observe Dirac crossings near the bulk A/X points close to the Fermi energy indicated by pink circles and arrows), although the crossing near X is situated slightly above $E_F$. The reason for the slight deviation of these Dirac crossings from the bulk high symmetry points A/X is that the 2D high symmetry points $\overline{\text{K}}'$ and $\overline{\text{K}}$ do not project exactly to A/X as can be seen in Fig. 1 (c). We also note the weakly-dispersive bands (highlighted by black arrows) located around $E~-~E_F$ = ~-~0.5 eV and $E~-~E_F$ = ~-~0.7 eV, which originate from the distorted kagome lattice.


\begin{figure*}
	\centering
	\includegraphics[width=\linewidth]{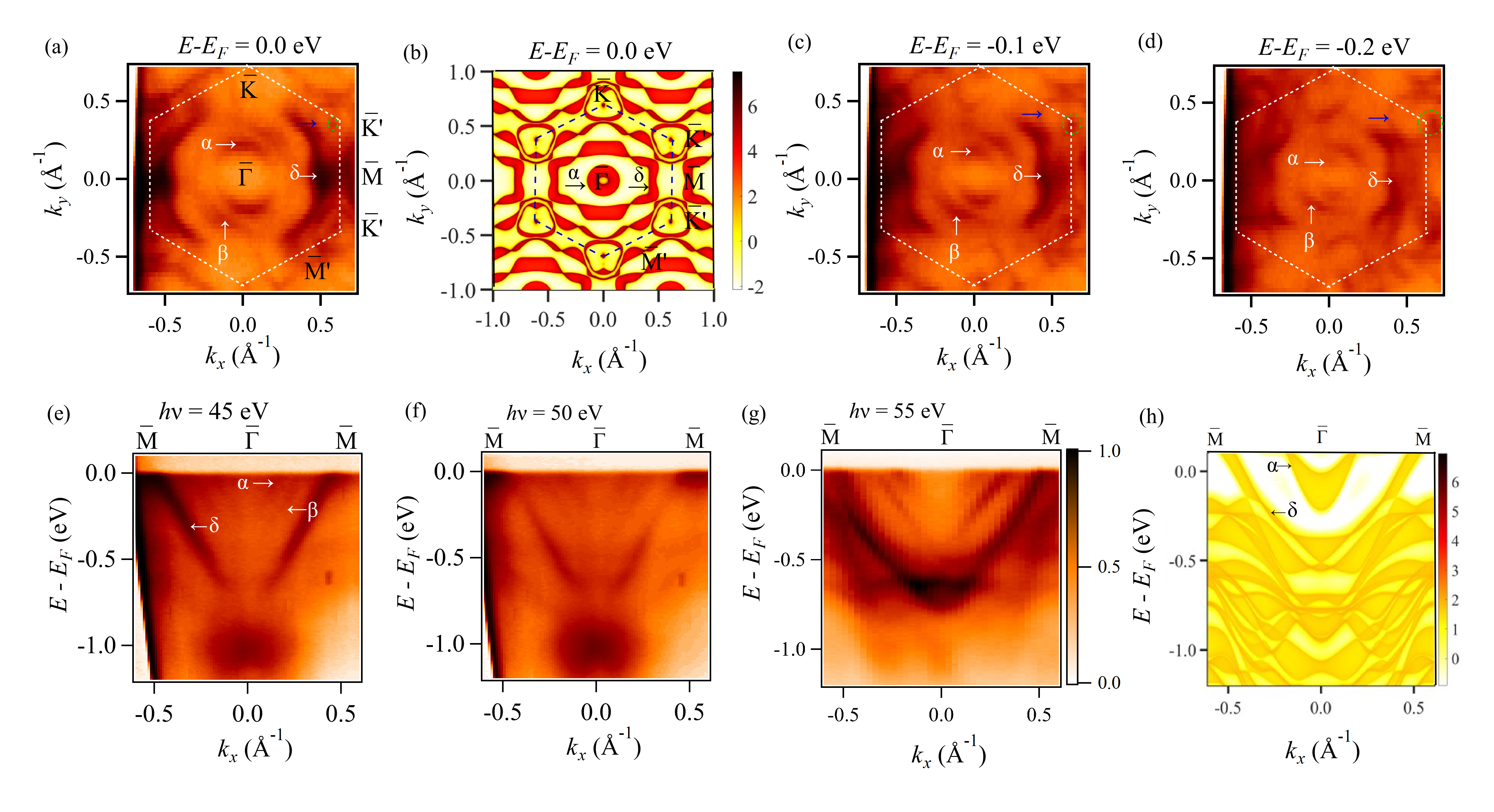}
	\caption{Band dispersion along the $\overline{\text{M}}$--$\overline{\Gamma}$--$\overline{\text{M}}$ high-symmetry line. (a) The measured Fermi surface (FS) with the surface-projected high-symmetry points indicated. (b) DFT calculated FS. (c-d) Constant energy contours (CECs) for binding energies of $E-E_F =$ -0.1 eV, and -0.2 eV, respectively. The ARPES-measured FS and CECs were obtained using an incident photon energy of $h\nu$ = 55 eV. (e-g) ARPES-measured band dispersions along the $\overline{\text{M}}-\overline{\Gamma}-\overline{\text{M}}$ high-symmetry lines taken using $h\nu =$ (e) 45 eV, (f) 50 eV, and (g) 55 eV incident photon energies, respectively. (h) Surface projected bulk band structure along the $\overline{\text{M}}-\overline{\Gamma}-\overline{\text{M}}$ high-symmetry line. ARPES measurements were performed at 15 K in the paramagnetic phase.  
	\label{fig:cecp}
    }
\end{figure*}

To corroborate the band dispersions predicted from DFT calculations, we have performed high-resolution ARPES measurements as presented in Figs. 2-4. In Figs. 2(a-d), we present the Fermi surface (FS) and constant energy contours (CECs) at various binding energies. Starting with the experimental FS shown in Fig. 2(a), we observe two circular-like pockets surrounding the $\overline{\Gamma}$ point, denoted as $\alpha$ and $ \beta$. Additionally, there is a large hexagonal pocket, indicated as $\delta$ in Fig. 2(a) along the $\overline{\Gamma}-\overline{\text{K}}(\overline{\text{K}}')-\overline{\text{M}}$ direction. Finally, there are a pair of small pockets centered at the $\overline{\text{K}}(\overline{\text{K}}')$ point. The inner pocket is nearly point-like, indicated by green small dotted circle, while the outer pocket features an apparent triangular shape (see blue arrow). The DFT calculated bulk FS, presented in Fig. 2(b) is in good agreement with our experimental results. The circular and hexagonal shape pockets at $\overline{\Gamma}$ high-symmetry point exhibit electron-like dispersion, evident from their decreasing volumes as binding energies increase indicated by white arrow from left to right in Figs. 2(c-d)). Conversely, both pockets at the $\overline{\text{K}}(\overline{\text{K}}')$ point (see blue arrow and small green circle) show an expansion in volume as binding energy rises, suggesting their hole-like characteristics; see supplementary material as well (Fig. S2).\\
\indent Figs. 2(e-g) presents the ARPES-measured band dispersion along the $\overline{\text{M}}-\overline{\Gamma}-\overline{\text{M}}$ direction. Firstly, two faint bands, denoted as $\alpha$ and $\beta$, and one intense band $\delta$ can be observed surrounding the $\overline{\Gamma}$ point, corresponding to the previously discussed circular and hexagonal like pockets. These $\alpha$ and $\beta$ bands become more prominent when increasing the photon energy from $h\nu =$ 45 eV (Fig. 2(e)) to $h\nu=$ 50 eV (Fig. 2(f)) and $h\nu = $ 55 eV (Fig. 2(g)). The innermost band, ($\alpha$) and the outermost band, ($\delta$) exhibit electron-like characteristics which is consistent with the FS and CECs shown in Figs. 2 (a, c, d). The second band ($\beta$), corresponding to the middle circular-like pocket, disperses further in binding energy, reaching what appears to be a flat band at around $E-E_F\approx$ -0.42 eV. Our bulk DFT calculations in Fig. 2 (h) are in good agreement with the results of ARPES measurements in Fig. 2 (e, f, g) reproducing the two electron-like pockets $\alpha$ and $\delta$, indicating their bulk origin. However, the band denoted as $\beta$ in the ARPES spectra is not reproduced in our bulk DFT calculations. When surface states are included in the calculations, see supplementary material (Fig. S7) shows the appearance of the electron-like pocket $\beta$ (black arrow), suggesting its surface origin \cite{SM}. To elucidate the topology of NdTi$_3$Bi$_4$, we computed the Z$_2$ invariant through tracking the Wannier charge center evolution and employing parity-based methods. The Z$_2$ number was calculated using the number of electrons as the number of occupied bands and we found that it is a weak topological insulator with Z$_2$=(0;011). 

\begin{figure*}
	\centering
	\includegraphics[width=\linewidth]{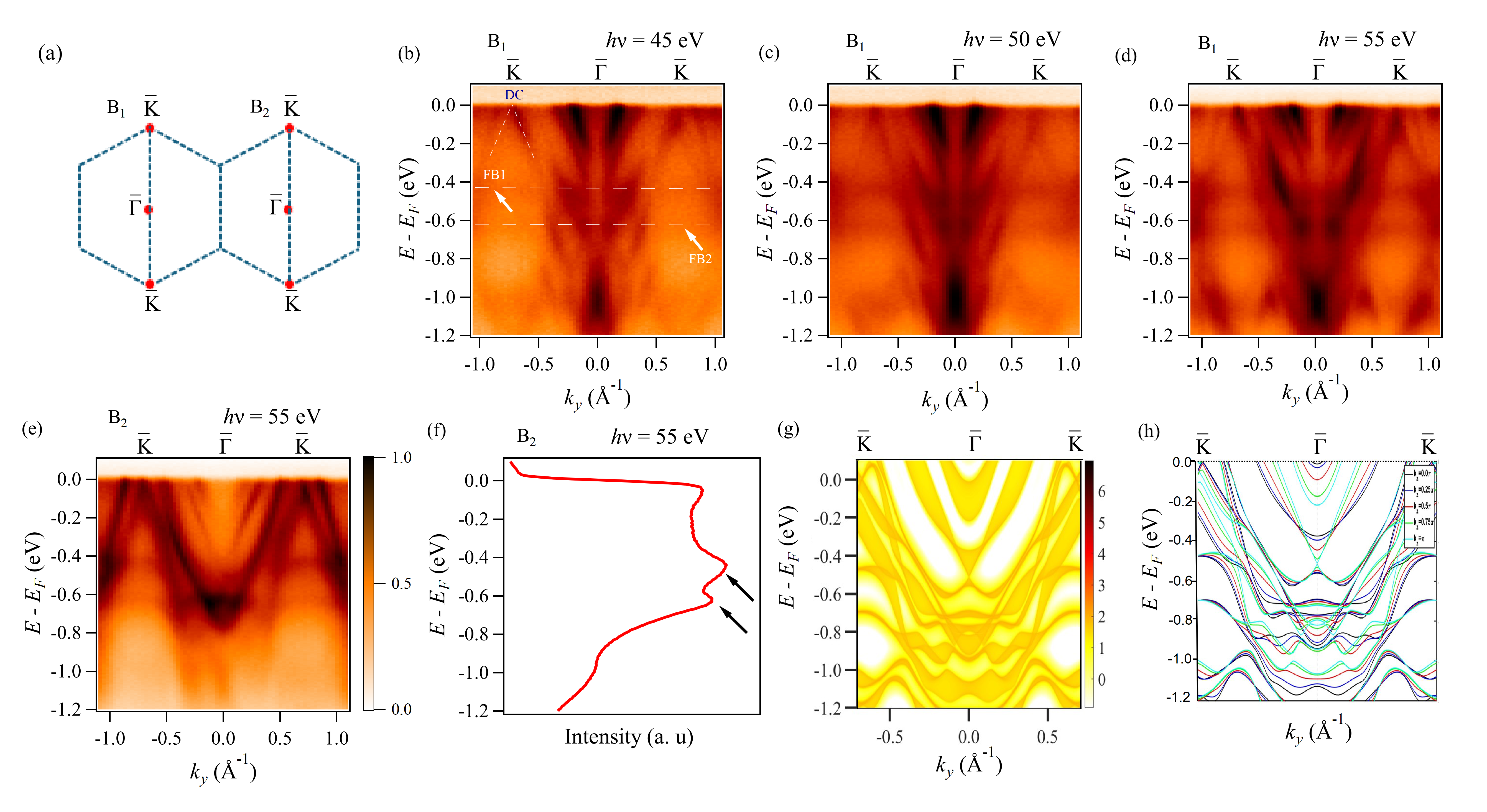}
	\caption{
    Observation of flat bands. (a) Schematic showing multiple brillouin zones (BZs). The $\overline{\text{K}}-\overline{\Gamma}-\overline{\text{K}}$ valence band dispersions are measured using (b) 45 eV, (c) 50 eV, and (d) 55 eV incident photon energies. Attention is drawn to the three relevant features, the linear crossings near $\overline{\text{K}}$, and the two flat bands located at $E-E_F$ = -0.42 eV and -0.63 eV through indications with white arrows in panel (b). (e, f) Band dispersion and EDC along $\overline{\text{K}}-\overline{\Gamma}-\overline{\text{K}}$ in the second BZ mentioned $B_{2}$ in (a). (g-h) Projected bulk band structure and 
\textit{k$_z$}-resolved bulk band structure along the
$\overline{\text{K}}-\overline{\Gamma}-\overline{\text{K}}$ high-symmetry lines. ARPES measurements were performed at 15 K in the paramagnetic phase.  
	\label{fig:dirac}
    }
\end{figure*}

Fig. 3(a) is a schematic showing multiple BZs. B$_1$ and B$_2$ denote cuts along the $\overline{\text{K}}-\overline{\Gamma}-\overline{\text{K}}$ direction in the first and second BZ, respectively. Figs. 3(b-e) exhibit dispersion cuts along the same direction for incident photon energies of 45 eV, 50 eV, and 55 eV, respectively. These dispersion cuts reveal the presence of two flat bands (indicated by white arrows and dashed lines in Fig. 3 (b)); see the Supplemental Material \cite{SM}; as well as linear Dirac bands around the $\overline{\text{K}}$ point, visible across all photon energies. Fig. 3(e) presents a cut along the B$_2$ direction, as marked in Fig. 3(a), taken in the second BZ, where the Dirac cone at $\overline{\text{K}}$ is more easily resolvable.
Additionally, the $\overline{\text{K}}-\overline{\Gamma}-\overline{\text{K}}$ cut reveals two flat bands at binding energies of approximately $E~-~E_F$ = -0.42 eV and $E~-~E_F$ = -0.63 eV, clearly evident in the energy distribution curve (EDC) illustrated in Fig. 3(f). This EDC analysis has been performed within the momentum range between -1 \AA$^{-1}$ to 1 \AA$^{-1}$ along the B$_2$ cut, as shown in Fig. 3(e). Orbital-projected DFT calculations, discussed previously in Fig. 1(d), attribute these flat bands to Ti \textit{d$_{xy}$} and Ti \textit{$d_{x^{2}-y^{2}}$} orbitals, suggesting their origin from the distorted kagome lattice arrangement formed by Ti atoms. 
Fig. 3(g) and Fig. 3(h) illustrate the projected bulk band structure and \textit{k$_z$}-resolved bulk band structure along the $\overline{\text{K}}-\overline{\Gamma}-\overline{\text{K}}$ high-symmetry lines, respectively. A comparative analysis suggests that the $\alpha$ and $\delta$ bands at the $\overline{\Gamma}$ point originates from the intersection of bulk bands at the \textit{k$_z$} = $\pi$ plane. These calculations, presented in Fig. 3(g) and Fig. 3(h), complement our ARPES findings. Additionally, surface-exclusive \textit{k$_z$}-resolved  DFT calculations, outlined in the Supplemental Material (Fig. S5) \cite{SM}, confirm this interpretation by showing no linear crossing at the $\overline{\Gamma}$-point.\\

\indent We now turn our attention towards elucidating the details of the VHSs previously discussed in Fig. 1(d) through DFT calculations. To validate these findings experimentally, we analyze sections along the high-symmetry line starting and ending by the red points in Fig. 4(a). Fig. 4(b) illustrates the band dispersion along this trajectory, from $\overline{\Gamma}$ to $\overline{\text{M}}$ to $\overline{\text{K}}’$ and back to $\overline{\Gamma}$. Of particular interest are the VHSs, identified as VHS1, VHS2, and VHS3 (indicated by white arrows), which emerge along the $\overline{\Gamma}$-$\overline{\text{M}}$-$\overline{\text{K}}’$-$\overline{\Gamma}$ high-symmetry line. Upon comparing with the DFT calculation depicted in Fig. 1(d), we observe a close correspondence between the theoretical predictions and experimental findings regarding the band behavior surrounding the VHSs.


\begin{figure}
	\includegraphics[width = 8.6 cm]{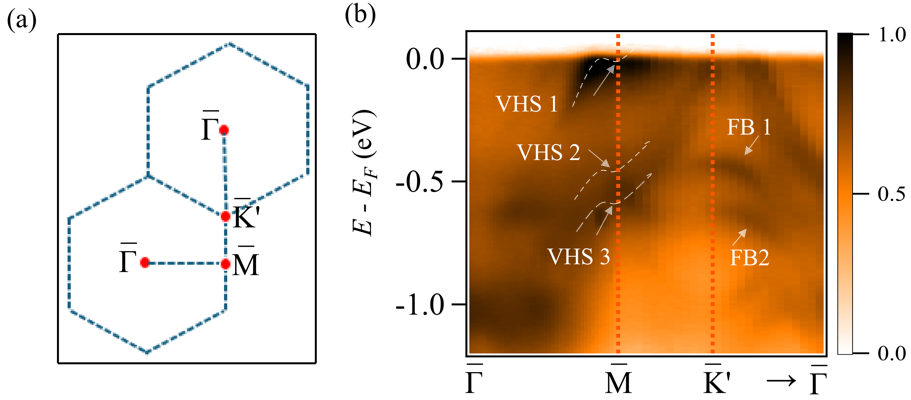}
	\caption{
    Observation of multiple Van hove singularities. 
    (a) Schematic showing multiple BZs demonstrating the direction of the cut presented in panel (b). (b) The band dispersion along the high-symmetry $\overline{\Gamma}-\overline{\text{M}}-\overline{\text{K}}'-\overline{\Gamma}$ path. ARPES measurements were performed at 15 K in the paramagnetic phase. 
	\label{fig:dirac}
    }
\end{figure}


In summary, we have performed ARPES measurements and DFT calculations for NdTi$_3$Bi$_4$, a newly identified distorted ferromagnetic kagome metal. Our ARPES measurements within the paramagnetic phase reveal the presence of multiple VHSs along with two flat bands. The ARPES measurements align well with our DFT calculations, indicating good agreement between experimental and theoretical results. Our DFT calculations suggest that these flat bands originate from Ti 3\textit{d$_{xy}$}/3\textit{$d_{x^{2}-y^{2}}$} orbitals, suggesting their origin from the destructive interference within the Ti kagome motif. We also observe Dirac cone at the $\overline{\text{K}}(\overline{\text{K}}')$ point. Our discovery of several VHSs and flat bands, along with non-trivial electronic topology reveal that NdTi$_3$Bi$_4$ will serve as a promising platform to explore the interplay of electronic correlation and topology in this new kagome metal.

\indent \textit{Note added.} During the review process of this manuscript, several related studies have been published. \cite{Chen, Guo134, Shi}.\\

M. N. acknowledges the support from the US Department of Energy (DOE), Office of Science, Basic Energy Sciences grant number DE-SC0024304 and the Air Force Office of Scientific Research MURI (Grant No. FA9550-20-1-0322).
Work performed by B.R.O. is sponsored by the Laboratory Directed Research and Development Program of Oak Ridge National Laboratory, managed by UT-Battelle, LLC, for the US Department of Energy. The work at Northeastern University was supported by the Air Force Office of Scientific Research under Award No. FA9550-20-1-0322, and it benefited from the computational resources of Northeastern University’s Advanced Scientific Computation Center (ASCC) and the Discovery Cluster. This research used resources of the Advanced Light Source, a U.S. Department of Energy Office of Science User Facility, under Contract No. DE-AC02-05CH11231. We thank Sung-Kwan Mo for beamline assistance at the Advanced Light Source (ALS), Lawrence Berkeley National Laboratory.

\def\bibsection{\section*{\refname}}

\vspace{2ex}

\raggedbottom
\clearpage
\pagebreak{}

\clearpage
\widetext
\begin{center}
\textbf{\large Supplemental Materials for: \\~\\\Large Observation of multiple flat bands and Van-Hove singularities in the distorted kagome metal NdTi$_3$Bi$_4$}\\
\end{center}
\setcounter{equation}{0}
\setcounter{figure}{0}
\setcounter{table}{0}
\setcounter{page}{1}
\makeatletter
\renewcommand{\theequation}{S\arabic{equation}}
\renewcommand{\figurename}{{Fig.}}
\renewcommand{\thefigure}{{{S\arabic{figure}}}}
\renewcommand{\bibnumfmt}[1]{[#1]}
\renewcommand{\citenumfont}[1]{#1}
\renewcommand{\tablename}{Supplementary Table}
\renewcommand{\thetable}{\arabic{table}}
\def\bibsection{\refname}
\renewcommand{\refname}{\noindent\textbf{Supplementary References}\\}

\begin{center}
\vspace{1 cm}
\textbf{SUPPLEMENTARY ARTICLES}   
\end{center}

\title{Supplemental Material:\\Observation of multiple flat bands and Van-Hove singularities in the distorted kagome metal NdTi$_3$Bi$_4$}
\author{Mazharul Islam Mondal}  \thanks{These authors contributed equally in this work.} \affiliation{Department of Physics, University of Central Florida, Orlando, Florida 32816, USA} 
\author{Anup Pradhan Sakhya}  \thanks{These authors contributed equally in this work.} \affiliation {Department of Physics, University of Central Florida, Orlando, Florida 32816, USA} 
\author{Milo Sprague} \affiliation{Department of Physics, University of Central Florida, Orlando, Florida 32816, USA} 
\author{Brenden R. Ortiz} \affiliation {Materials Science and Technology Division, Oak Ridge National Laboratory, Oak Ridge, Tennessee 37830, USA}
\author{Matthew Matzelle} \affiliation{Department of Physics, Northeastern University, Boston, Massachusetts 02115, USA}
\author{Arun K Kumay} \affiliation{Department of Physics, University of Central Florida, Orlando, Florida 32816, USA} 
\author{Avike Seal} \affiliation{Department of Physics, University of Central Florida, Orlando, Florida 32816, USA} 
\author{Barun Ghosh} \affiliation{Department of Physics, Northeastern University, Boston, Massachusetts 02115, USA}
\author{Arun Bansil} \affiliation{Department of Physics, Northeastern University, Boston, Massachusetts 02115, USA}
\author{Madhab Neupane} \thanks{Corresponding author:\href{mailto:madhab.neupane@ucf.edu}{madhab.neupane@ucf.edu}}\affiliation{Department of Physics, University of Central Florida, Orlando, Florida 32816, USA}

\maketitle

\date{\today}

\noindent\textbf{1. Methods}\\
\noindent\textbf{1.1 Experimental and Computational details.}\\
NdTi$_3$Bi$_4$ single crystals were grown utilizing a bismuth self-flux method. Nd (Alfa 99.9\%), Ti(Alfa 99.9\% powder), and Bi (Alfa 99.999\% low-oxide shot) elemental reagents were mixed at a ratio of 2:3:12 and placed in 2 mL Canfield crucibles with a catch crucible and a porous frit. The crucibles were airtight under approximately 0.7 atm of argon gas within fused silica ampoules. Each composition was gradually heated to 1050 \textdegree C at a rate of 200 \textdegree C/hr. After reaching 1050 \textdegree C, the samples were allowed to thermalize and homogenize for 12-18 hours before cooling to 500 \textdegree C at a rate of 2 \textdegree C/hr. Excess bismuth was eliminated through centrifugation at 500 \textdegree C.\\

\noindent First principles calculations were done using projector augmented-wave pseudopotentials \cite{Kresse} within the Vienna ab initio simulation package (VASP) \cite{Kresse1, Kresse2, Kresse3, Kresse4}. The strongly constrained and appropriately normed (SCAN) meta-GGA exchange-correlation functional \cite{Sun} was used. An energy cutoff of 500 eV was adopted for the plane wave basis and the Brillouin Zone was sampled by an 8$\times$8$\times$8  $\Gamma$-centered grid. The electronic structure was self-consistent until the energy difference between electronic steps was less than 10$^{-6}$ eV. The effects of spin-orbit coupling are included self-consistently.\\

\noindent\textbf{1.2. Synchrotron measurements.}\\
The electronic structure of NdTi$_{3}$Bi$_{4}$ was measured using synchrotron-based ARPES at the Advanced Light Source (ALS) on beamline 10.0.1.1. Single crystals of NdTi$_{3}$Bi$_{4}$ were cut into small pieces and affixed onto a copper post using silver epoxy. The sample preparation was conducted within a glove box to prevent oxidation. The measurements were conducted with a Scienta R4000 hemispherical electron analyzer. To ensure a clean surface, the samples were cleaved in an ultra-high vacuum environment (pressure of $5\times10^{-11}$ Torr). The synchrotron measurements had an energy resolution better than $20$~meV and an angular resolution finer than $0.2^{\circ}$. The stability of NdTi$_{3}$Bi$_{4}$ sample was excellent during the typical 20-hour measurement period with no signs of degradation. The crystals were prepared as small, flat shiny pieces, mounted on copper posts, and ceramic posts were attached on top using silver epoxy. These prepared samples were then loaded into the measurement setup for studying the electronic structure.\\

\noindent\textbf{1. Honeycomb and triangular lattice.}\\
In Fig. S1 (a, b), we present the honeycomb-like and triangular lattice formed by Bi atoms.\\

\noindent\textbf{2. Fermi surface.}\\
In Fig. S2 (a, b), we present the Fermi surface (FS) map measured along the (001) direction using a photon energy of 60 eV. This FS covered multiple of brillouin zones (BZs) indicated by white dashed lines and mention various high symmetry points. We have included annotations to explicitly mark the key features around the $K$ and ${K}'$ points. The inner small circular pocket is indicated by a green dotted circle and the outer triangular shape pocket is indicated by blue arrow (see Fig. S2 (b). The log colors of the FS plot has been added in the Fig. S2 (c) so that we can resolve the band features more clearly.\\ 

\noindent\textbf{3. ARPES measured band dispersion along the $\overline{\text{M}}$--$\overline{\Gamma}$--$\overline{\text{M}}$} direction\\
Schematic of multiple Brillouin zones are shown in the figure. Fig. S3 (a) where we indicate the ARPES-measured band dispersion cut (C$_1$) along the $\overline{\text{M}}$--$\overline{\Gamma}$--$\overline{\text{M}}$ high-symmetry lines. In Fig. S3 (b, c), we present the dispersion cut along the $\overline{\text{M}}$--$\overline{\Gamma}$--$\overline{\text{M}}$ direction and its second derivative where the bands, $\alpha$, $\beta$, and $\delta$ at the $\overline{\Gamma}$ point is distinctly resolved.\\

\noindent\textbf{4. ARPES measured band dispersion along the $\overline{\text{K}}$--$\overline{\Gamma}$--$\overline{\text{K}}$--$\overline{\text{M}}$--$\overline{\text{K}}$ high-symmetry lines.}\\
Fig. S4 (a) illustrates the band dispersion as measured by ARPES along the $\overline{\text{K}}$--$\overline{\Gamma}$--$\overline{\text{K}}$--$\overline{\text{M}}$--$\overline{\text{K}}$ high-symmetry lines. The measurement was conducted at a photon energy of 55 eV. Two flat bands are discernible at binding energies approximately $\sim$ -0.42 eV and $\sim$ -0.63 eV. The energy distribution curves (EDCs) depicted in Fig. S4(b) validate the presence of two peaks corresponding to these flat bands.\\

\begin{figure*}
	\includegraphics[width=14cm]{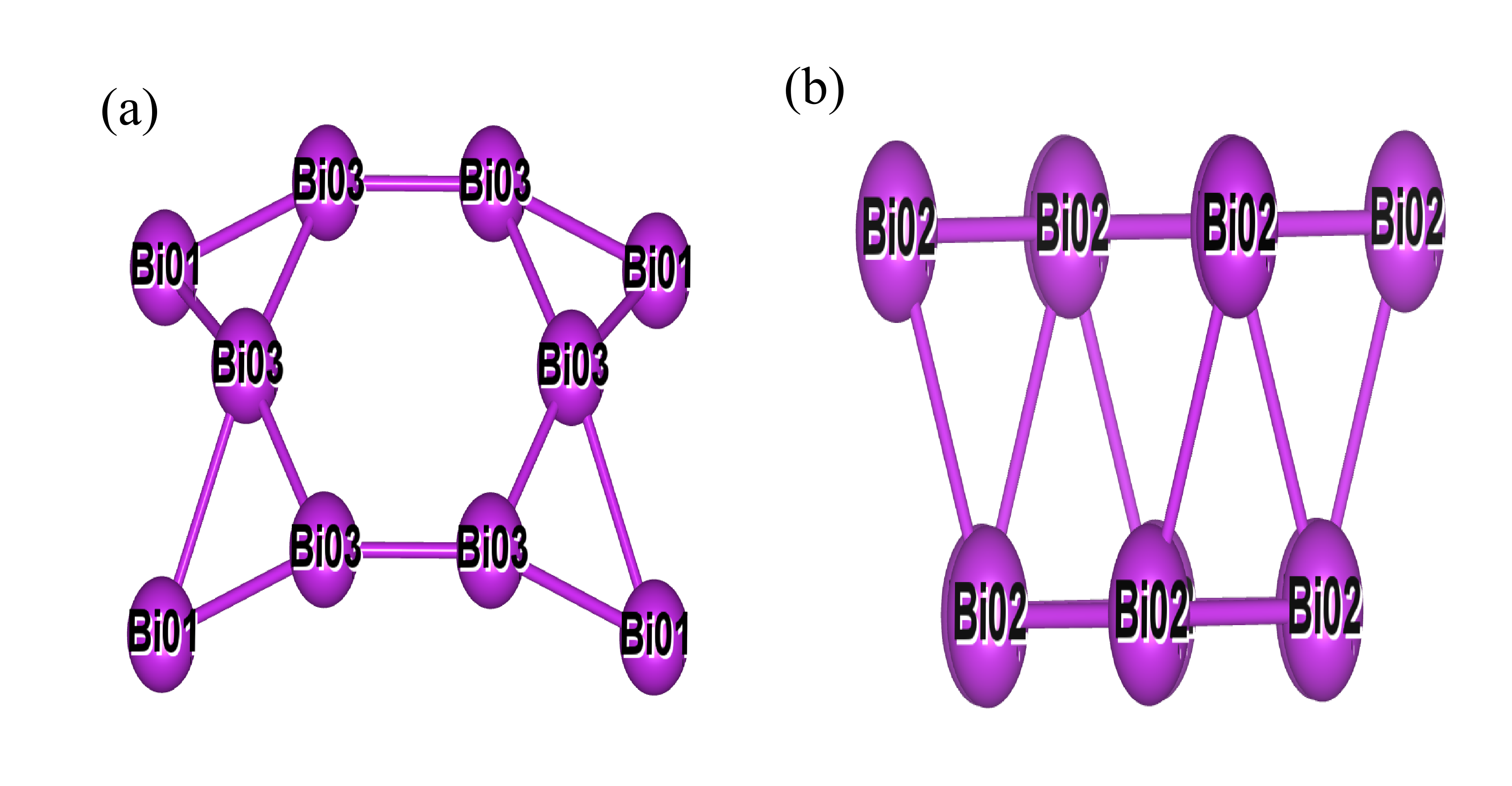}
\centering
\caption{ (a, b) Shows the honeycomb lattice and triangular lattice formed by Bi atoms which are accordingly marked in the figures.}
\end{figure*}

\begin{figure*}
	\includegraphics[width=14cm]{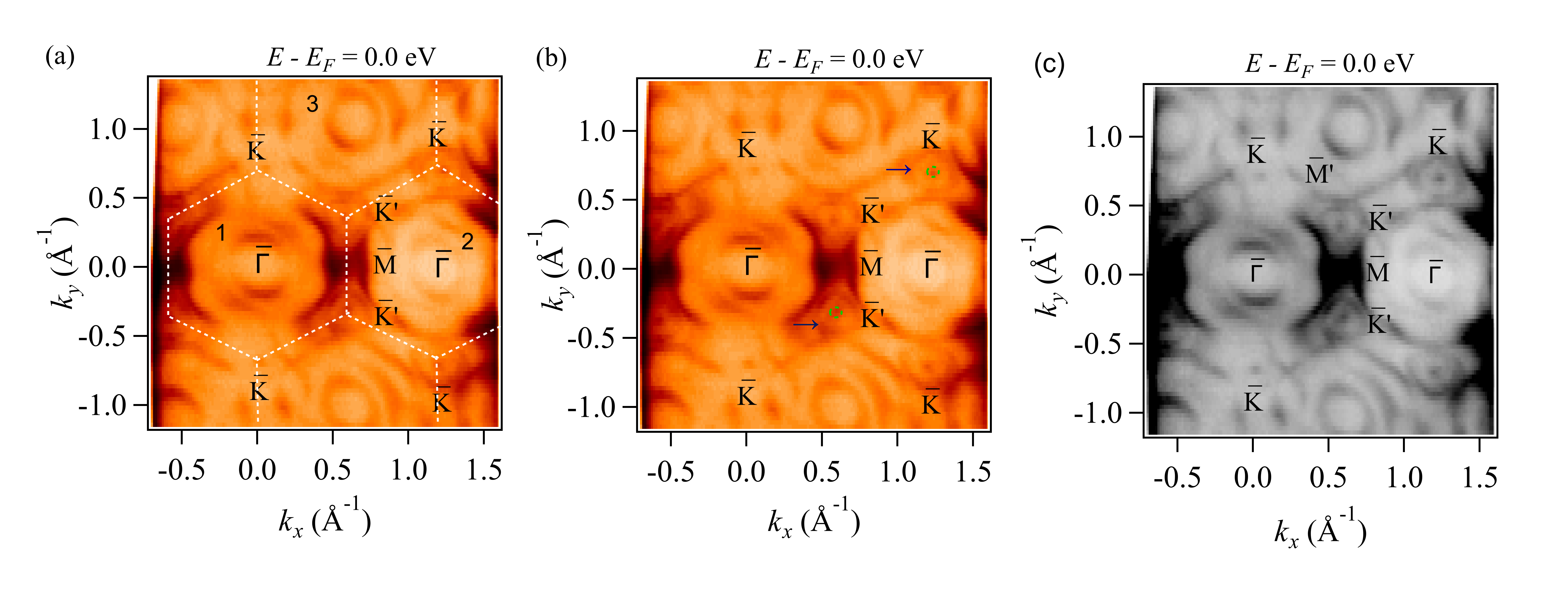}
\centering
\caption{ Experimental Fermi surface (FS) measured along the (001) direction at an incident photon energy of 60 eV, (a) with appending the BZs (b) without appending the BZs but indicating the band features at high symmetry $\overline{\text{K}}$ and $\overline{\text{K}}'$ points. (c) FS in log colors. The measurements were conducted at the ALS beamline 10.0.1.1 in the paramagnetic phase.}
\end{figure*}

\begin{figure*}
	\includegraphics[width=14cm]{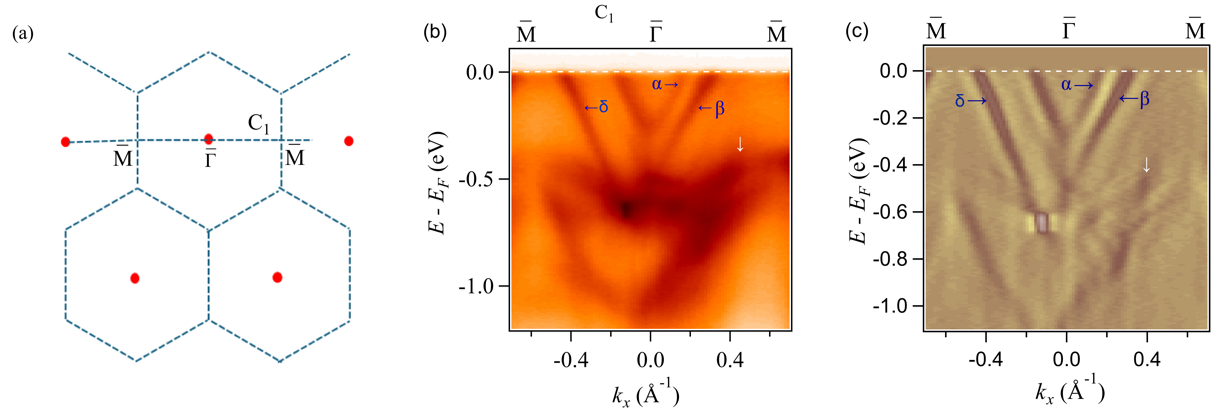}
\centering
\caption{ Wide Fermi surface (FS) map and electronic band structure. (a) Schematic of the FS map with various high-symmetry points indicated. (b) Band structure along the $\overline{\text{M}}$--$\overline{\Gamma}$--$\overline{\text{M}}$ high symmetry lines. (c) Second derivative of $\overline{\text{M}}$--$\overline{\Gamma}$--$\overline{\text{M}}$ dispersion cut. ARPES measurements were performed at 15 K in the paramagnetic phase.}
\end{figure*}

\begin{figure*}
\centering
\includegraphics[width=14cm]{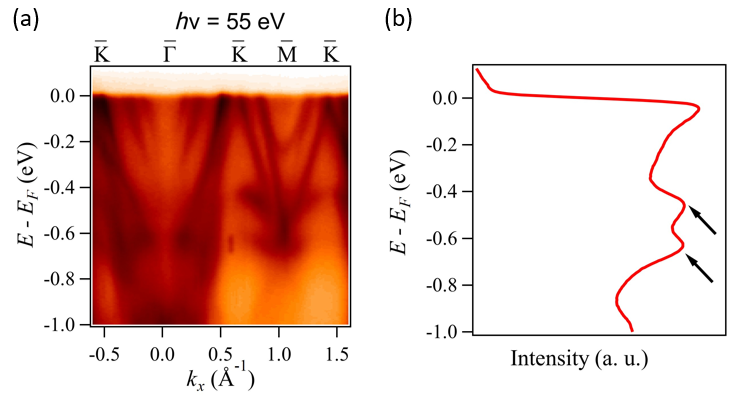}
\caption{(a) Band dispersion measured along the $\overline{\text{K}}$--$\overline{\Gamma}$--$\overline{\text{K}}$--$\overline{\text{M}}$--$\overline{\text{K}}$ high-symmetry lines using 55 eV photon energy. (b) EDCs along the high-symmetry lines as depicted in (a). ARPES measurements were performed at 15 K in the paramagnetic phase.}
\end{figure*}

\begin{figure*}
\centering
\includegraphics[width=10cm]{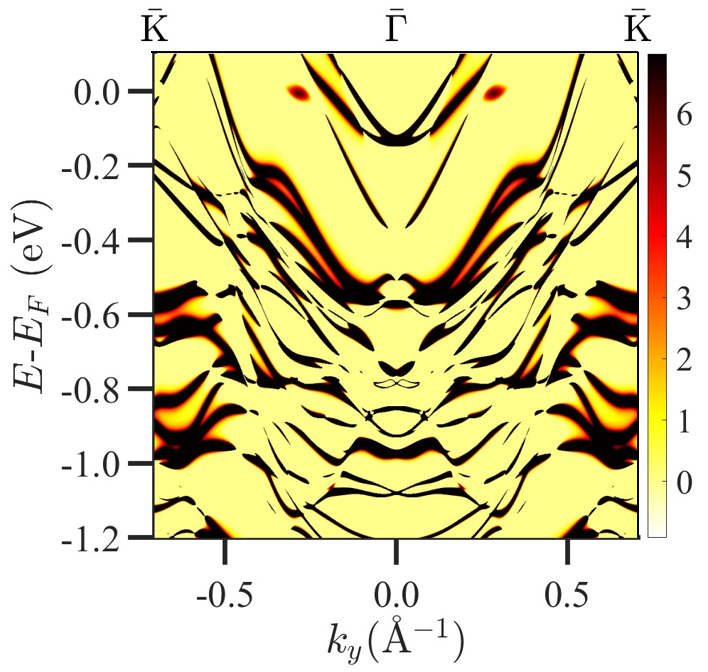}
\caption{Calculated surface band structure along the $\overline{\text{K}}$--$\overline{\Gamma}$--$\overline{\text{K}}$ high-symmetry line.}
\end{figure*}

\noindent\textbf{5. Calculated surface state dispersion along the $\overline{\text{K}}$--$\overline{\Gamma}$--$\overline{\text{K}}$ high-symmetry line.}\\
Fig. 3 (g, h) in the main text presents the projected bulk bands and the \textit{k$_z$}-resolved bulk bands, respectively, along the $\overline{\text{K}}$--$\overline{\Gamma}$--$\overline{\text{K}}$ high-symmetry line. A linear dispersing state is observed within the bulk spectrum at the \textit{k$_z$} = $\pi$ plane. For comparison, the complementary surface bands along the $\overline{\text{K}}$--$\overline{\Gamma}$--$\overline{\text{K}}$ high-symmetry line are presented in Fig. S5. Notably, there is a clear absence of a linear crossing at the $\overline{\Gamma}$-point in these surface bands.\\

\noindent\textbf{6. Electronic dispersion along the $\overline{\text{K}}$--$\overline{\Gamma}$--$\overline{\text{K}}’$ direction.}\\
Fig. S6 (a, b) we have shown the FS that covers multiple Bzs and dispersion cut along the $\overline{\text{K}}$--$\overline{\Gamma}$--$\overline{\text{K}}’$ in order to observe the difference of electronic band structure between $\overline{\text{K}}$ and $\overline{\text{K}}'$ direction. Finally, we can resolve the mild changes of band structure where the flat band become slightly dispersive along the $\overline{\Gamma}$--$\overline{\text{K}}'$ direction and Dirac cone is formed slightly at a higher binding energy to $\overline{\Gamma}$--$\overline{\text{K}}'$.\\

\noindent\textbf{7. Calculated surface state dispersion along the $\overline{\text{M}}$--$\overline{\Gamma}$--$\overline{\text{M}}$ high-symmetry line.}\\
The $\beta$ band is not resolved in our bulk DFT calculations. To investigate further, we performed  DFT calculated Fermi surface that includes the surface state and a bulk-projected surface band structure along the $\overline{\text{M}}$--$\overline{\Gamma}$--$\overline{\text{M}}$ high-symmetry direction (see Fig. S7 (a, b)), which reveals the presence of the $\beta$ band (indicated by a black arrow), suggesting its surface origin. To bolster the evidence of the surface nature of the $\beta$ band, we analysis the photon-energy dependence of $\alpha$, $\beta$ and $\delta$ pockets. We consider the MDCs peak position of $\alpha$, $\beta$ and $\delta$ bands along the $\overline{\text{M}}$--$\overline{\Gamma}$--$\overline{\text{M}}$ high-symmetry direction at fermi energy where We observe that the peak position of $\alpha$ and $\delta$ bands show some photon-energy dependence, whereas the $\beta$ peak remains at the same position (see Fig. S7 (c)), which indicates the surface nature of the $\beta$ band.\\  

\noindent\textbf{8. Density of state calculation (DOS).}\\
As to the role of the Nd chains in the observed anisotropy, the DFT results show very little contribution of the Nd states near the Fermi level. This is apparent in Fig. S8 which shows the density of states (DOS) decomposed by elements near the Fermi level. Ti is the only element that has a strong contribution to the states near the Fermi level suggesting its sole role in the elongation of the Fermi surface sheet.\\

\begin{figure*}
\centering
\includegraphics[width=14cm]{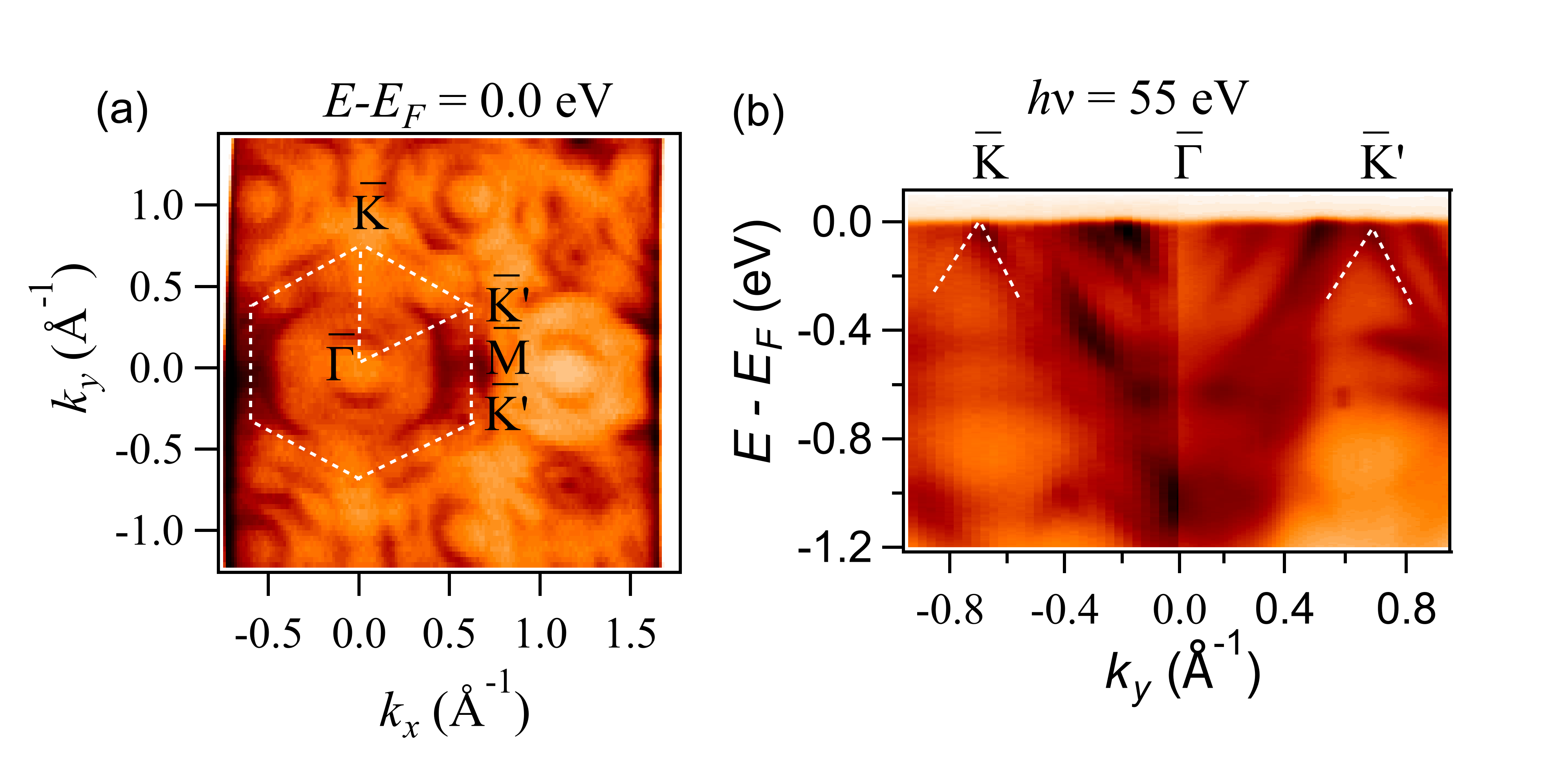}
\caption{(a) Experimental Fermi surface showing multiple brillouin zones. (b) Electronic band structure along the $\overline{\text{K}}$--$\overline{\Gamma}$--$\overline{\text{K}}'$ high-symmetry direction.}
\end{figure*}

\begin{figure*}
\centering
\includegraphics[width=15cm]{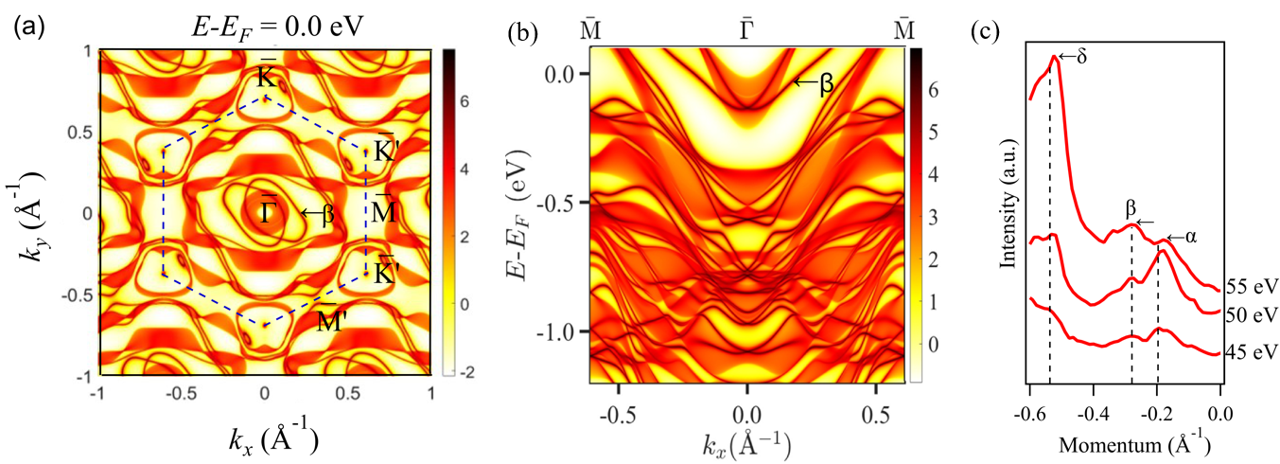}
\caption{ (a) DFT calculated Fermi surface includes the surface state. (b) Bulk projected surface band structure along the $\overline{\text{M}}$--$\overline{\Gamma}$--$\overline{\text{M}}$ high-symmetry direction. (c) Momentum distribution curves (MDCs) stacked over different incident photon energies (45 eV, 50 eV, 55 eV) that cover multiple K$_{z}$ planes. MDCs taken along the $\overline{\text{M}}$--$\overline{\Gamma}$--$\overline{\text{M}}$ high-symmetry direction at Fermi energy. Black arrows indicate the pick position of $\alpha$, $\beta$ and $\delta$ bands.}
\end{figure*}

\begin{figure*}
\centering
\includegraphics[width=15cm]{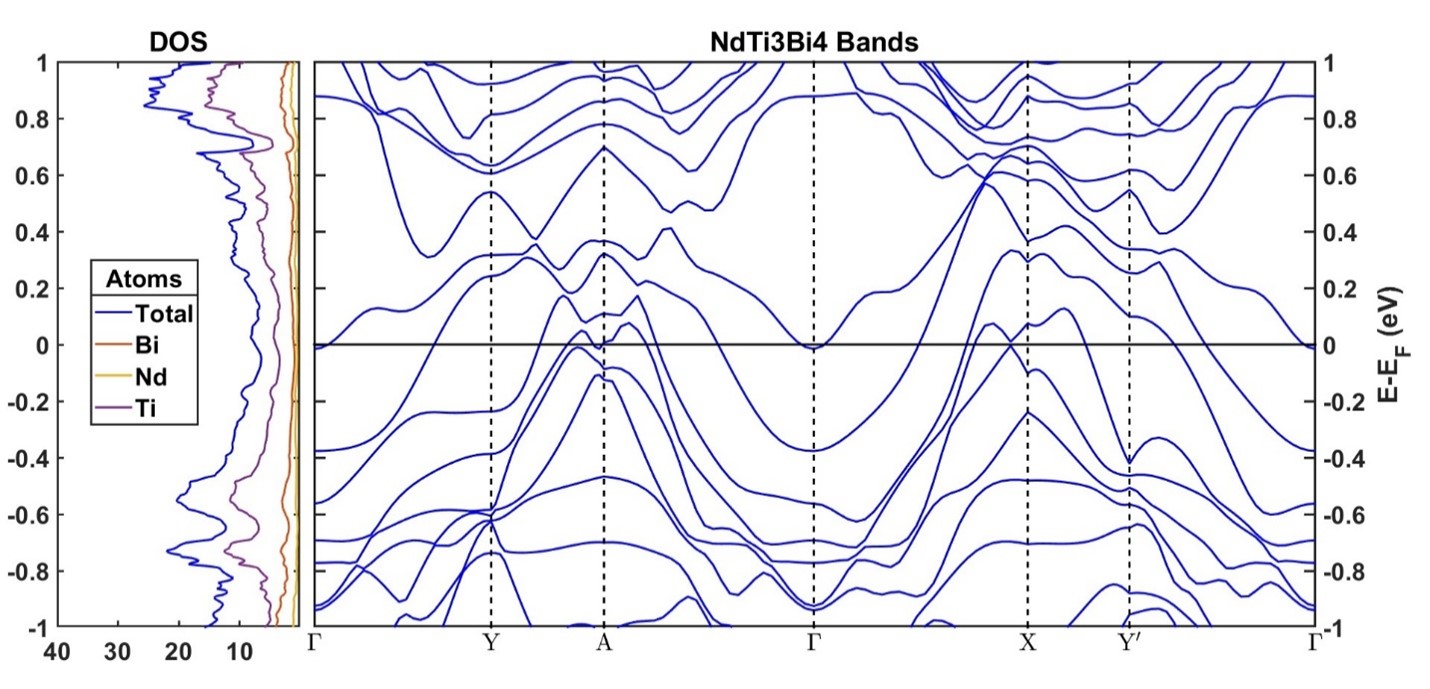}
\caption{: DOS and band structure of NdTi3Bi4. Note the weak Nd contribution to the states at the Fermi level.}
\end{figure*}

\clearpage

\end{document}